\documentclass[
    ,final            
  ]
  {aipproc}

\layoutstyle{6x9}

\begin{document}

\title{Dark Energy and Neutrino Model in SUSY}


\author{Ryo Takahashi}{
  address={Graduate School of  Science and Technology,
 Niigata University,  950-2181 Niigata, Japan}
}\renewcommand{\thefootnote}{\fnsymbol{footnote}}
\footnote[1]{talked at the SUSY06, the 14th International Conference
on Supersymmetry and the Unification of Fundamental Interactions, UC
Irvine, California, 12-17 June 2006.} 

\author{Morimitsu Tanimoto}{
  address={Department of Physics,
Niigata University,  950-2181 Niigata, Japan}
}

\begin{abstract}
We discuss the effect of  the supersymmetry breaking 
on the Mass Varying Neutrinos(MaVaNs) scenario. Especially, the
effect mediated by the gravitational interaction between the hidden
sector and the dark energy sector is studied. A model including a
chiral superfield in the dark sector and the right-handed neutrino
superfield is proposed. Evolutions of the neutrino mass and the
equation of state parameter are presented in the model. 
\end{abstract}

\maketitle


\section{Introduction}
In a dynamical dark energy model proposed by Fardon, Nelson and Weiner
(MaVaNs), relic neutrinos could form a negative pressure fluid and
cause the cosmic acceleration \cite{Weiner}. In this model, an unknown
scalar field which is called ``acceleron'' is introduced and neutrinos
are assumed to interact through a new scalar force. The acceleron sits
at the instantaneous minimum of its potential, and the cosmic
expansion only modulates this minimum through changes in the neutrino
density. Therefore, the neutrino mass depends on its number density
and changes with the evolution of the Universe. The equation of state
parameter $w$ and the dark energy density also evolve with the
neutrino mass. Those evolutions depend on a model of the scalar
potential and the relation between the acceleron and the neutrino mass
strongly. Typical examples of the potential and some supersymmetric
models have been discussed in ref. \cite{Peccei,our,fardon,our2}. 

In this talk, we present a model including the supersymmetry breaking
effect mediated by the gravity. Then we show evolutions of the
neutrino mass and the equation of state parameter in the model.

\section{Supersymmetric MaVaNs}

The basic assumption of the MaVaNs with supersymmetry is to
introduce a chiral superfield $A$ in the dark sector, which is a
singlet under the gauge group of the standard model. We assume that
the superfield $A$ couples to both the left-handed lepton doublet
superfield $L$ and the right-handed neutrino superfield $R$. 

In this framework, we suppose the superpotential
 \begin{equation}
  W=\frac{\lambda }{6}A^3+\frac{M_A}{2}AA+m_DLA+M_DLR
    +\frac{M_R}{2}RR,\label{W}
 \end{equation}
where $M_A$, $M_D$, $M_R$ and $m_D$ are mass parameters. The scalar
 and spinor component of $A$ are $(\phi ,\psi )$, and the scalar
 component is assumed to be the acceleron which cause the present
 cosmic acceleration. The spinor component is a sterile neutrino. 

In the MaVaNs scenario, the dark energy is assumed to be the sum of
the neutrino energy density and the scalar potential for the acceleron:
 $\rho _{\mbox{{\tiny DE}}}=\rho _\nu +V(\phi )$. The effective scalar
potential, the lagrangian density and the effective neutrino mass
matrix are given as 
 \begin{eqnarray}
  V(\phi )&=&\frac{\lambda ^2}{4}|\phi |^4+M_A^2|\phi |^2
           +m_D^2|\phi |^2,\\
  \mathcal{L}&=&\lambda \phi\psi\psi +M_A\psi\psi
                +m_D\bar{\nu}_L\psi+M_D\bar{\nu}_L\nu _R
                +M_R\nu _R\nu _R+h.c.,\\ 
  \label{lag}
  \mathcal{M}&\simeq&
   \left(
   \begin{array}{cc}
    c   & m_D \\
    m_D & M_A+\lambda\phi
   \end{array}
  \right).\label{MM}
 \end{eqnarray}
The neutrino mass matrix is written in the basis of $(\bar{\nu}_L,\psi
   )$, where we assume $\lambda\phi\ll M_D\ll M_R$ and define $c\equiv
   -M_D^2/M_R$. The first term of the $(1,1)$ element of this matrix
   corresponds to the usual term given by the seesaw mechanism. It is
   remarked that only the mass of a sterile neutrino is variable in
   the case of the vanishing mixing ($m_D=0$) between the left-handed
   and a sterile neutrino on cosmological time scale. The finite
   mixing ($m_D\neq 0$) makes the mass of the left-handed neutrino
   variable. 

In the MaVaNs scenario, there are two constraints on the scalar
potential. The first one comes from observations of the Universe,
which is that the magnitude of the present dark energy density is
about $0.74\rho _c$, $\rho _c$ being the critical density. Thus, the
first constraint turns to $V(\phi ^0)=0.74\rho _c-\rho _\nu
^0$. ``$0$'' represents a value at the present epoch.

The second one is the stationary condition. In this scenario, the
neutrino is a dynamical quantity which is a function of the
acceleron. Therefore, the dark energy density should be stationary
with respect to the variation of the neutrino mass:
  $\partial\rho _{\mbox{{\tiny DE}}}/\partial m_\nu=0$.
This equation is rewritten by using the cosmic temperature $T$:
$\partial V/\partial m_\nu =-T^3\partial F/\partial\xi$, where
$\xi\equiv m_\nu /T$, $\rho _\nu =T^4F(\xi )$ and $F(\xi
)\equiv\frac{1}{\pi ^2}\int _0^\infty dyy^2\sqrt{y^2+\xi ^2}/(e^y+1)$.
We can get the time evolution of the neutrino mass from this
stationary condition. Since the stationary condition should be
always satisfied in the evolution of the Universe, this one at the
present epoch is the second constraint on the scalar potential:
 \begin{eqnarray}
  \left.\frac{\partial V(\phi )}{\partial m_\nu}\right|
  _{m_\nu =m_\nu^0}
  =\left.-T^3\frac{\partial F(\xi )}{\partial\xi}\right|
   _{m_\nu =m_\nu^0,T=T_0}.
  \label{stationary2}
 \end{eqnarray} 
In addition to two constraints for the potential, we also have two
relations between the acceleron and the neutrino mass:
 \begin{equation}
  m_i=\frac{c+M_A+\lambda\phi}{2}
              \pm\frac{\sqrt{[c-(M_A+\lambda\phi )]^2+4m_D^2}}
                      {2},
  \label{nmass}
 \end{equation}
where the plus and the minus sign correspond to the left-handed and a
sterile neutrino mass, respectively ($i=\nu _L$, $\psi$).

Next, we will consider the dynamics of the acceleron field. In order
that the acceleron does not vary significantly on distance of
inter-neutrino spacing, the acceleron mass at the present epoch must
be less than $\mathcal{O}(10^{-4}\mbox{eV})$ \cite{Weiner}. Here and
below, we fix the present acceleron mass as $m_\phi ^0=10^{-4}\mbox
{ eV}$. Once we adjust parameters which satisfy some conditions for
the MaVaNs model, we can have evolutions of the neutrino mass from the
stationary condition.  

The dark energy is characterized by the evolution of the equation of
state parameter $w$. The equation of state in this scenario is derived
from the energy conservation equation in the Robertson-Walker
background and the stationary condition: $w+1=[4-h(\xi )]\rho _\nu
/(3\rho _{\mbox{{\tiny DE}}})$, where $h(\xi )\equiv\xi\frac{\partial
F(\xi )}{\partial\xi}/F(\xi )$. It seems that $w$ in this scenario
depend on the neutrino mass and the cosmic temperature. This means
that $w$ varies with the evolution of the Universe unlike the
cosmological constant.

\section{Effect of supersymmetry breaking}
In order to consider the effect of supersymmetry breaking in the dark
sector, we assume a superfield $X$, which breaks supersymmetry, in the 
hidden sector, and the chiral superfield $A$ in the dark sector is
assumed to interact with the hidden sector only through the
gravity. Once supersymmetry is broken at TeV scale, its effect is
transmitted to the dark sector through the operators $\int d^4\theta
X^\dagger XA^\dagger A/ M_{p\ell}^2$ and $\int d^4\theta (X^\dagger
+X)A^\dagger A/M_{p\ell}$. $M_{p\ell}$ is the Planck mass. Then, the
scale of the soft terms
$F_X(\mbox{TeV}^2)/M_{p\ell}\sim\mathcal{O}(10^{-3}$-$10^{-2}\mbox{eV})$
is expected. Such a framework was discussed in the ``acceleressence''
scenario \cite{acceleressence}. 

In this framework, taking supersymmetry breaking effect into account,
the scalar potential is given by
 \begin{equation}
  V(\phi )=\frac{\lambda ^2}{4}|\phi |^4
           -\frac{\kappa}{3}(\phi ^3+h.c.)+M_A^2|\phi |^2
           +m_D^2|\phi |^2-m^2|\phi |^2+V_0,
 \label{V1}
 \end{equation}
where $\kappa$ and $m$ are supersymmetry breaking parameters, and $V_0$ is 
a constant determined by the condition that the cosmological constant
is vanishing at the true minimum of the acceleron potential. This
scalar potential is the same one presented in \cite{acceleressence}. 

When the mixing between the left-handed and a sterile neutrino is
vanishing, $m_D=0$ in the neutrino mass matrix (\ref{MM}). Then we
have the mass of the left-handed and a sterile neutrino as $m_{\nu _L}
=c$ and  $m_\psi =M_A+\lambda\phi$, respectively.
In this case, we find that only the mass of a sterile neutrino is
variable on cosmological time scale.





In the case of the finite mixing between the
left-handed and a sterile neutrino ($m_D\neq 0$), the
left-handed and a sterile neutrino mass are given by Eq. (\ref{nmass}).
We can expect that both the mass of the left-handed and a sterile
neutrino have varied on cosmological time scale due to the term of the 
acceleron dependence.

Taking typical values for four parameters as $\lambda =1$,
$m_D=10^{-3}\mbox{ eV}$, $m_{\nu _L}^0=2\times 10^{-2}\mbox{ eV}$ and
$m_\psi ^0=10^{-2}\mbox{ eV}$, we have $\phi ^0\simeq -1.31\times
10^{-5}\mbox{ eV}$, $c\simeq 1.99\times 10^{-2}\mbox{ eV}$, $M_A\simeq
1.01\times 10^{-2}\mbox{ eV}$, $m\simeq 1.02\times 10^{-2}\mbox{ eV}$
and $\kappa\simeq 4.34\times 10^{-3}\mbox{ eV}$.
\begin{figure}[t]
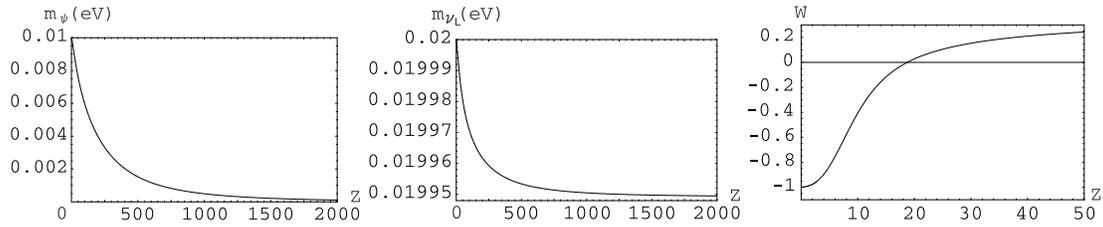

\includegraphics[height=.13\textheight]{fig4.ai}
\includegraphics[height=.13\textheight]{fig5.ai}
\includegraphics[height=.13\textheight]{fig7.ai}
\caption{Evolutions of a sterile and the left-handed neutrino mass and
the equation of state parameter. $z$ is a red shift parameter.}
\label{fig:3}
\end{figure}
Evolutions of the left-handed and a sterile neutrino mass and the
equation of state parameter are shown in Fig. \ref{fig:3}. The mass of
the left-handed neutrino is variable unlike the vanishing mixing
case. The mixing does not almost affect evolutions of a sterile
neutrino mass and the equation of state parameter.

\section{Summary}
We presented a supersymmetric MaVaNs model including the effects of
the supersymmetry breaking mediated by the gravity. Evolutions of the
neutrino mass and the equation of state parameter have been calculated
in the model. Our model has a chiral superfield in the dark sector,
whose scalar component causes the present cosmic acceleration, and the
right-handed neutrino superfield. In our framework, supersymmetry is
broken in the hidden sector at TeV scale and the effect is assumed to
be transmitted to the  dark sector only through the gravity. Then, the 
scale of soft parameters are
$\mathcal{O}(10^{-3}$-$10^{-2})(\mbox{eV})$ is expected. 

In our model, the mixing between the left-handed and a sterile
neutrino makes the left-handed neutrino mass variable. The mixing does
not almost affect evolutions of the sterile neutrino mass and the
equation of state. 

\bibliographystyle{aipproc}   


\end{document}